\pdfoutput=1
\documentclass[a4paper]{jpconf}
\usepackage{citesort}
\usepackage{graphicx}
\usepackage{wrapfig,rotating}
\usepackage{epsfig}
\usepackage{cite}
\usepackage{color}
\usepackage{booktabs}

\begin{document}
\title{Data Preservation in High Energy Physics}

\author{Roman Kogler, David M. South, Michael Steder \\ on behalf of the ICFA DPHEP Study Group}

\address{Deutsches Elektronen Synchrotron, Notkestra\ss e 85, 22607 Hamburg, Germany}

\ead{roman.kogler@desy.de, david.south@desy.de, michael.steder@desy.de}

\begin{abstract}

  Data from high-energy physics experiments are collected with
  significant financial and human effort and are mostly unique.
  However, until recently no coherent strategy existed for data
  preservation and re-use, and many important and complex data sets
  have simply been lost. While the current focus is on the LHC at
  CERN, in the current period several important and unique
  experimental programs at other facilities are coming to an end,
  including those at HERA, b-factories and the Tevatron. To address
  this issue, an inter-experimental study group on HEP data
  preservation and long-term analysis (DPHEP) was convened at the end
  of 2008. The group now aims to publish a full and detailed review of
  the present status of data preservation in high energy physics. This
  contribution summarises the results of the DPHEP study group,
  describing the challenges of data preservation in high energy
  physics and the group's first conclusions and recommendations. The
  physics motivation for data preservation, generic computing and
  preservation models, technological expectations and governance
  aspects at local and international levels are examined.

\end{abstract}

\section{Introduction}
\label{sec:intro}

Particle physics experiments are designed to probe the structure of
matter, the character of fundamental interactions and to ultimately
extend our understanding of nature.
Since the advent of collider experiments the available range in energy
and intensity has been enlarged by many orders of magnitude
\cite{panofsky}.
However, the development, building and commissioning of colliders and
the experimental detectors takes considerable human, technological and
financial effort.
Most colliders and their associated scientific programmes have been
unique in energy range, process dynamics or experimental techniques.
The data preservation effort aims to ensure long-term availability of
these data after the end of the experimental collaborations.


Nowadays the focus of the particle physics community is on the Large
Hadron Collider (LHC) at CERN, which operates mainly as a $pp$
collider, currently at a centre--of--mass energy of $7$~TeV.
At the same time, a generation of other high energy physics (HEP)
experiments have concluded their data taking and the experimental
collaborations are finishing their physics programmes.
These include the HERA, LEP, KEK and PEP
experiments~\cite{southchep10}, as well as those at the Tevatron,
where data taking has recently ended~\cite{tevatronshutdown}.
Data collected from these experiments continue to be crucial to our
understanding of particle physics, ranging from precision measurements
\cite{jade, jade2} to searches for new signatures beyond the Standard
Model~\cite{alephhiggs}.
Besides the ongoing analyses that remain to be completed, these data
may also provide important future scientific
opportunities~\cite{bethkeqcd10}.
Moreover, the data from these experiments are often unique in terms of
the initial state particles and are unlikely to be superseded anytime
soon, even considering such future projects as the LHeC~\cite{lhec}.
The issue of HEP data becoming inaccessible due to technological
advances in operating systems, storage hardware or data loss becomes
imminent already at the LHC, where the lower--energy and
lower--luminosity data collected at $0.9$, $2.36$ and $7$~TeV in 2010
and 2011 will not be repeated.
Such datasets, which are unique in terms of energy regime and
experimental conditions should be secured in order to ensure re-use
and future accessibility~\cite{cmsdp}.
It is therefore prudent to envisage some form of conservation of the
respective data sets. However, HEP has little or no tradition or clear
current model of long term preservation of data in a meaningful and
useful way. The preservation of and supported long term access to the
data is generally not part of the planning, software design or budget
of a HEP experiment~\cite{tevatrondp}.


In the following, the physics case for data preservation is examined,
followed by an overview of the study group on Data Preservation and
Long Term Analysis in High Energy Physics (DPHEP). Different aspects
of HEP data are reviewed and a summary is given of the different
models for data preservation identified by the DPHEP study group.
Current inter--experimental data preservation initiatives are then
presented, followed by some words on governance and structures, before
finally concluding with an outlook and summary of future working
directions.

\section{The Physics Case for Data Preservation}
\label{sec:physcase}

\begin{figure}[t]
  \centering
  \hspace{-0.25cm}
  \begin{minipage}{0.27\textwidth}
    \vspace{-0.03cm}
    \includegraphics[width=1.1\textwidth]{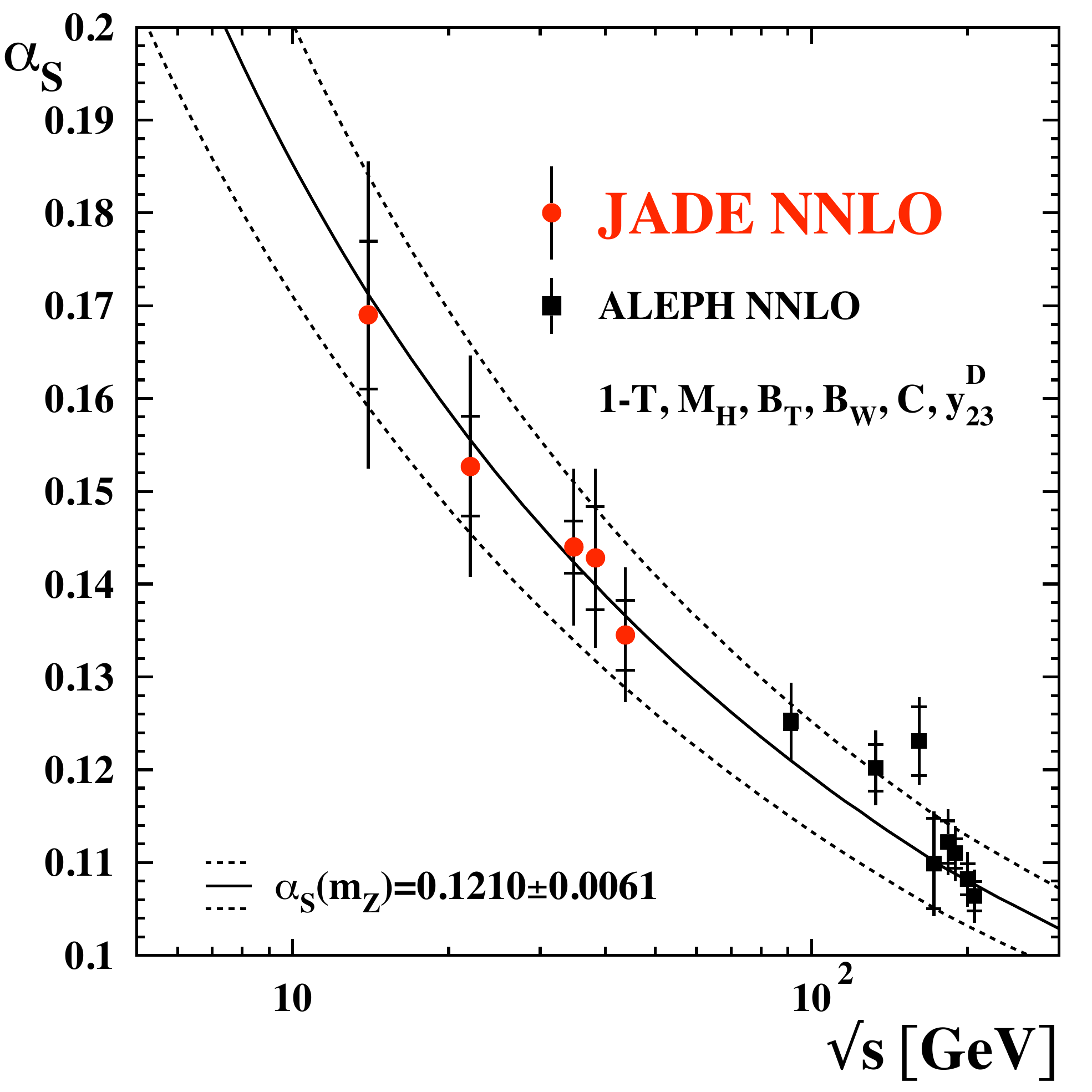}
  \end{minipage}
  \hspace{2pc}
  \begin{minipage}{0.35\textwidth}
    \vspace{-0.25cm}
    \hspace{-0.8cm}
    \includegraphics[width=1.225\textwidth]{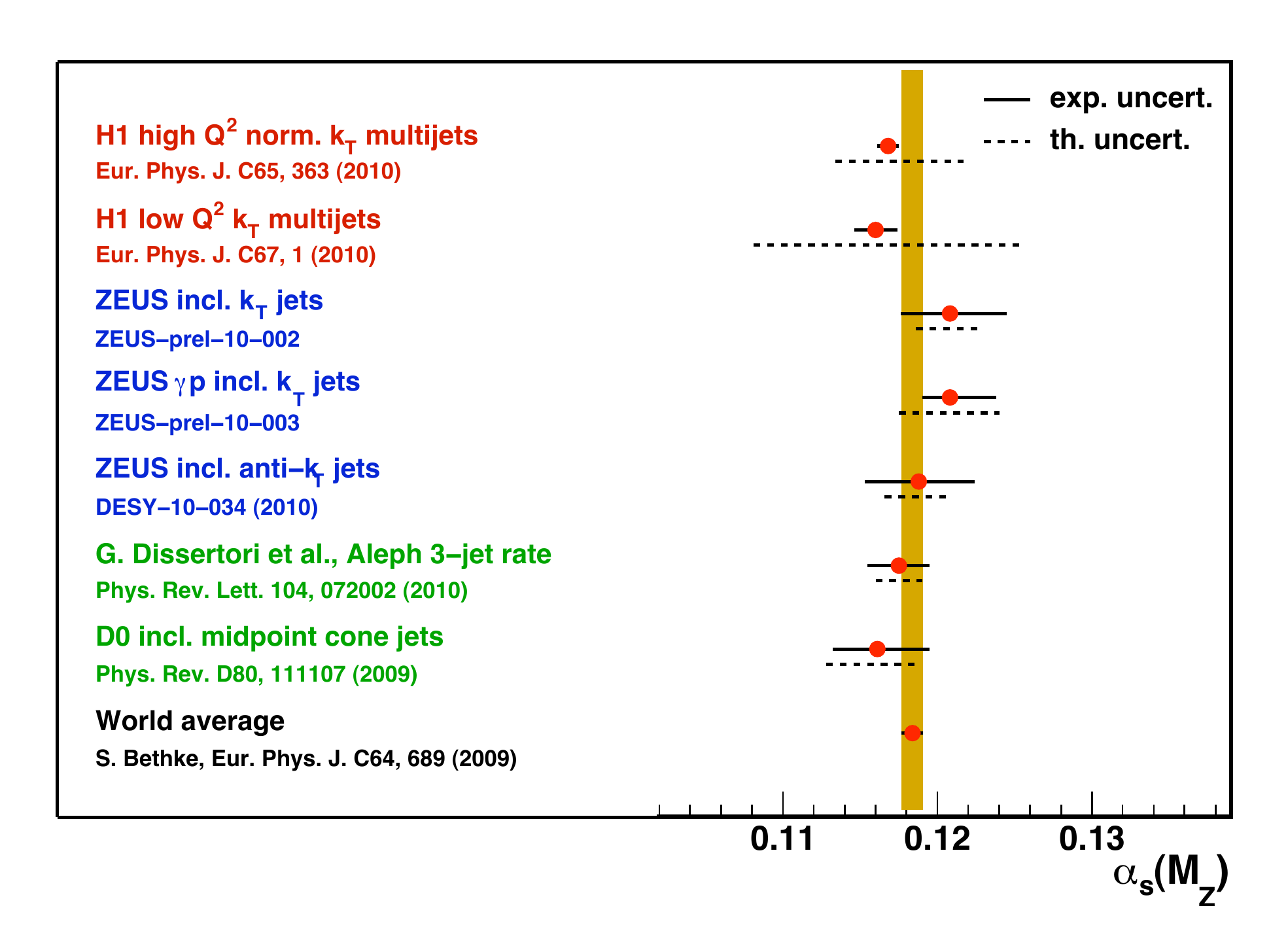}
  \end{minipage}
  \hspace{0.22cm}
  \begin{minipage}{0.29\textwidth}
    \vspace{-0.585cm}
    \includegraphics[width=1.121\textwidth]{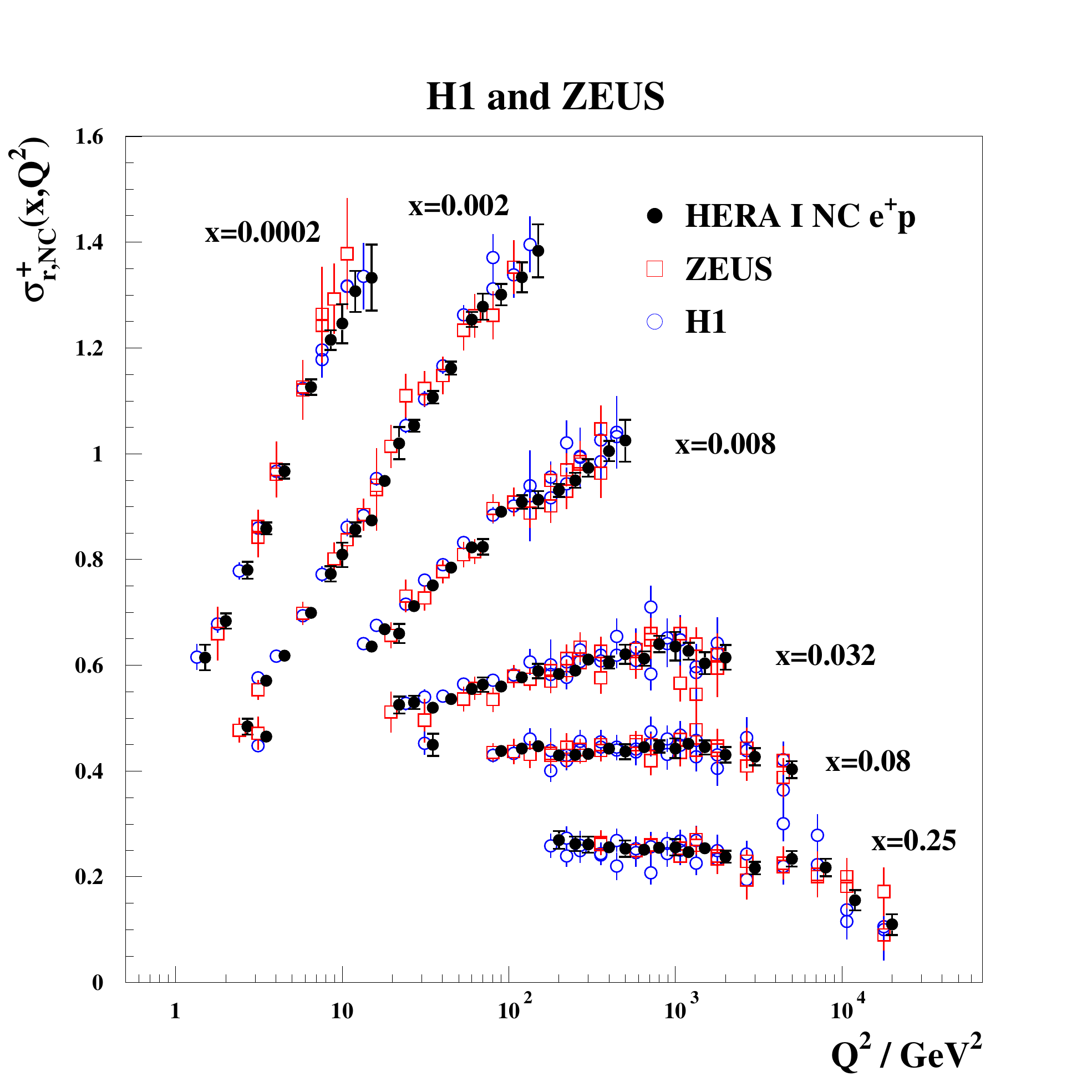}
  \end{minipage} 
  \caption{\label{fig:physcase} Left: Measurements of the strong
    coupling, $\alpha _{s}$ from an event shape analysis of JADE data
    at various centre--of--mass energies, $\sqrt{s}$. The full and
    dashed lines indicate the result from the JADE NNLO
    analysis~\cite{jade2}; Middle: Recent determinations of the strong
    coupling $\alpha_{s}(M_{Z})$ from a variety of experiments
    compared to the 2009 world average~\cite{kogler}; Right: HERA
    combined NC $e^+p$ reduced cross sections as a function of $Q^2$
    for different $x$ bins~\cite{f2_hera}.}
\end{figure}

The motivation behind data preservation in HEP should have its roots
in physics.
A detailed report on physics cases for data preservation is given
elsewhere \cite{dpheppub1}, only a brief review is presented in these
proceedings.
One of the main assumptions concerning experimental HEP data is that
older data will always be superseded by that from the next generation
experiment, usually at the next energy frontier.
However, this is not always the case and several scenarios exist where
the preservation of experimental HEP data would be of benefit to the
particle physics community:
\begin{itemize}

\item An extension of the existing physics programme may be necessary
  to ensure the long term completion of ongoing analysis.
  In particular, precision analyses continue long after the end of
  data taking, making use of the full statistical power and the best
  knowledge of the systematic uncertainties.
  This scenario has proven to be true for LEP~\cite{travistail} and a
  similar tail in the publication timeline is expected for H1 and
  BaBar~\cite{babar}.

\item It may be favourable to re-do previous measurements to achieve
  an increased precision via new and improved theoretical calculations
  (MC models) or newly developed analysis techniques.
  A re-analysis of the JADE (1979--1986) data has lead to a
  significant improvement in the determination of the strong coupling
  $\alpha_{s}(M_{Z})$, as shown in figure~\ref{fig:physcase}~(left),
  in an energy range that is still unique~\cite{jade,jade2}.
  Figure~\ref{fig:physcase}~(middle) shows a variety of
  $\alpha_{s}(M_{Z})$ measurements, as well as the current world
  average, where it can be seen that for the latest H1 measurements
  the theoretical uncertainty dominates the error~\cite{kogler}.
  In a situation that mirrors this JADE analysis, it is hoped that the
  uncertainty on $\alpha_{s}(M_{Z})$ will be further reduced at some
  point in the future by re--analysing the very accurate HERA data
  once improved theoretical predictions become available.

\item Combining similar measurements from different experiments
  increases the statistical significance and reduces systematic
  uncertainties via cross--calibration techniques to arrive at a more
  precise result.
  In figure~\ref{fig:physcase}~(right) individual and combined H1 and
  ZEUS measurements of the reduced neutral current cross section are
  shown.
  The combined cross sections serve as the sole input for a QCD
  analysis at NLO which determines a new set of parton distributions,
  HERAPDF1.0, with reduced experimental uncertainties~\cite{f2_hera}.

\item If new phenomena are found by analysing data recorded at the LHC
  or some other future collider, it may be useful or even mandatory to
  go back, if possible, and verify such results using older data.

\end{itemize}

\section{DPHEP}
\label{sec:DPHEP}

\begin{figure}[t]
  \begin{center}
    \includegraphics[width=0.8\textwidth, clip=true, trim=4cm 5cm 0cm 6cm]{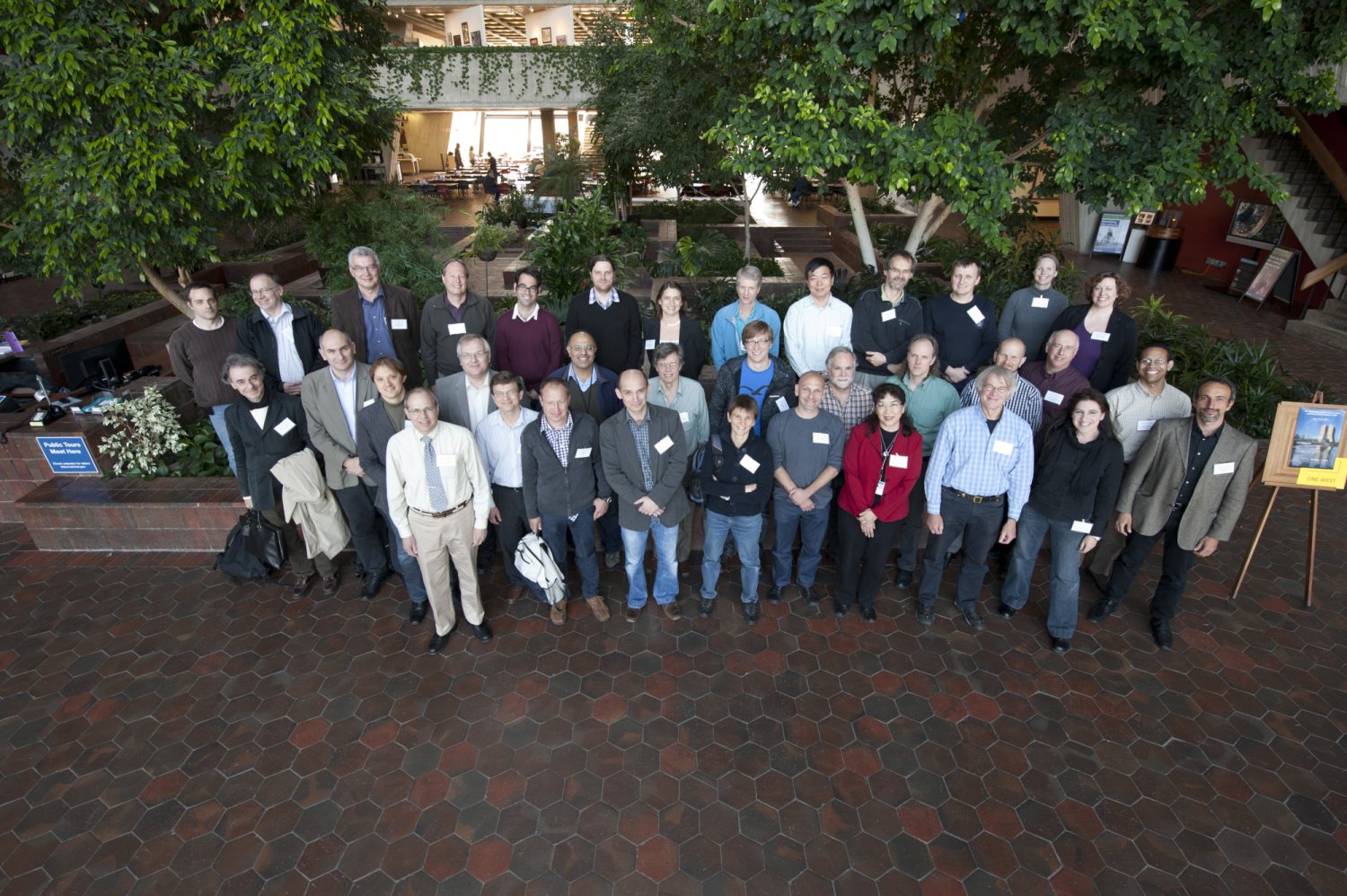}
  \end{center}
  \vspace{-0.3cm}
 \caption{\label{fig:people} Participants of the fifth DPHEP workshop at Fermilab, May 2011.}
 \vspace{-0.3cm}
\end{figure}

To address the issue of data preservation in a systematic way, the
DPHEP study group was formed at the end of $2008$~\cite{dpheporg}.
The aims of the study group include to confront the data models,
clarify the concepts, set a common language and investigate the
technical aspects of data preservation in HEP.
The experiments BaBar, Belle, BES-III, CLAS, CLEO, CDF, D{\O}, H1 and
ZEUS are all represented in DPHEP, with representatives from the LHC
experiments ALICE, ATLAS, CMS and LHCb having joined the study group
recently. The associated computing centres at CERN
(Switzerland/France), DESY (Germany), Fermilab (USA), IHEP (China),
JLAB (USA), KEK (Japan) and SLAC (USA) are all represented in DPHEP.
A series of workshops~\cite{dpheporg} have taken place over the last
three years, beginning at DESY in January 2009 and most recently at
Fermilab in May 2011.
The study group is officially endorsed with a mandate by the
International Committee for Future Accelerators (ICFA)~\cite{icfa} and
the first DPHEP recommendations were published in $2009$, summarising
the initial findings and setting out future working
directions~\cite{dpheporg}.
The role of the DPHEP study group is to provide international
coordination of data preservation efforts in high energy physics and
to provide a set of recommendations for past, present and future HEP
experiments.
Participants of the fifth DPHEP workshop at Fermilab are shown in
figure~\ref{fig:people}.

\section{What is HEP data?}
\label{sec:HEPdata}
In order to identify different models of data preservation, it is
necessary to survey the different aspects of HEP data which contain
important information for physics analyses.


The collision data recorded by the experiments are usually structured
in different layers, each successive layer containing decreasing
complexity as the detector based quantities such as measured charges
or track hits are consecutively replaced by reconstructed and
calibrated tracks, particle candidates and jets.
Hence the top--layer data have the full potential for physics
analyses, but improvements for example of the reconstruction
algorithms or calibrations can only be incorporated if the first
layer, typically referred to as the raw data, is available, thus
retaining full flexibility.
However, since these data usually have a total volume of a few
PB\footnote{The collisions recorded by the LHC experiments result in
  $10$'s of TB of data per day, or up to $15$~PB per year.}, today's
computing centres are able to store them without significant
additional costs.


In addition to the raw and reconstructed data the various software,
such as simulation, reconstruction and analysis software need to be
considered.
If the experimental software is not available the possibility to study
new observables or to incorporate new reconstruction algorithms,
detector simulations or event generators is lost.
Without a well defined and understood software environment the
scientific potential of the data may be limited.


In every experimental collaboration different levels of documentation
are available. Publications of data analyses or detector studies may
be in journals, on SPIRES or on arXiv, with the physics results
usually stored in a database like HEPDATA~\cite{hepdata}.
Detailed information about the analyses is usually only available in
Ph.D.\ or masters' theses or internal notes, where usually not all of
them are electronically available.
Many types of internal meta--data may also exist, such as the detailed
detector layout and performance, hardware replacements, manuals or the
documentation of meetings.


The implementation of a data preservation model as early as possible
in the lifetime of an experiment may greatly increase the chance that
the data will be available in the long term, and may also simplify the
data analysis in the final years of the collaboration.


The unique expertise of collaboration members is also at risk, as the
person power associated to the experiment decreases. By planning a
transition of the collaboration structure to something more suited to
an archival mode, this particular loss may be minimised (see section
\ref{sec:governance}).

\section{Models of Data Preservation}
\label{sec:models}
The resurrection of the JADE analysis chain to perform the analyses
described above, proved to be an eventful exercise and often a subject
of luck rather than careful planning~\cite{bethkeqcd10}.
The general status of the LEP data, which was recorded as recently as
the year $2000$, is a concern, despite the continued paper output. A
recent review of the status of the data of the four experiments
identified that efforts are needed to ensure long term
access~\cite{lep}.


The different data preservation models established by DPHEP are
summarised in table~\ref{tab:models}, organised in levels of
increasing benefit, which comes with increasing complexity and cost.
Each level is associated with use cases, and the preservation model
adopted by an experiment should reflect the level of analysis expected
to be available in the future. More details on each of the
preservation levels is given in the first DPHEP
publication~\cite{dpheppub1}.
\begin{table}[tb]
  \begin{center}
    \renewcommand{\arraystretch}{1.2} 
    \begin{tabular}{ p{0.2cm} p{7.0cm} p{7.0cm}}
      \toprule
      \multicolumn{2}{l}{Preservation Model}  & Use Case \\
      \toprule
      1. & Provide additional documentation &  Publication-related information search\\
      \midrule
      2. & Preserve the data in a simplified format & Outreach, simple training analyses\\
      \midrule
      3. & Preserve the analysis level software and the data format & Full scientific analysis based on existing reconstruction\\
      \midrule
      4. & Preserve the reconstruction and simulation software and basic level data & Full potential of the experimental data\\
      \bottomrule
    \end{tabular}
  \end{center}
  \vspace{-0.35cm}
  \caption{\label{tab:models} Various data preservation models, listed in order of increasing complexity, cost and benefits.}
\end{table}


Although a level $1$ preservation model, to provide additional
documentation, is considered the simplest, this still requires some,
often substantial, activity by the experiment.
The HERA collaborations, as well as BaBar, are all currently involved
in dedicated efforts to safeguard and streamline the available
documentation concerning their respective experiments (see section
\ref{sec:inspire}).
A level $2$ preservation, to conserve the experimental data in
simplified format, is considered to be unsuitable for high level
analyses, lacking the depth to allow, for example, detailed systematic
studies to be performed.
However, such a format is ideal for education and outreach purposes,
which many experiments in the study group are also actively interested
in (see section \ref{sec:outreach}).
Past experiences with old HEP data indicate that new analyses and
complete re--analyses are only possible if all the necessary
ingredients to retrieve, reconstruct and understand the data are
accounted for.
Only with the full flexibility does the full potential of the data
remain, equivalent to the DPHEP level $4$ data preservation.
Accordingly, the majority of participating experiments in the study
group plan for a level $4$ preservation programme, although different
approaches are employed concerning how this goal can be achieved.

\section{Common Data Preservation Projects}
\label{sec:projects}
Since the formation of DPHEP, activities and models of the experiments
have aligned to a certain degree and joint initiatives have been
launched, related to all four data preservation levels as described in
the following.

\subsection{A generic validation suite}
\label{sec:validation}

\begin{wrapfigure}{r}{5cm}
  \vspace{-0.4cm}
  \centering
  \includegraphics[width=5cm]{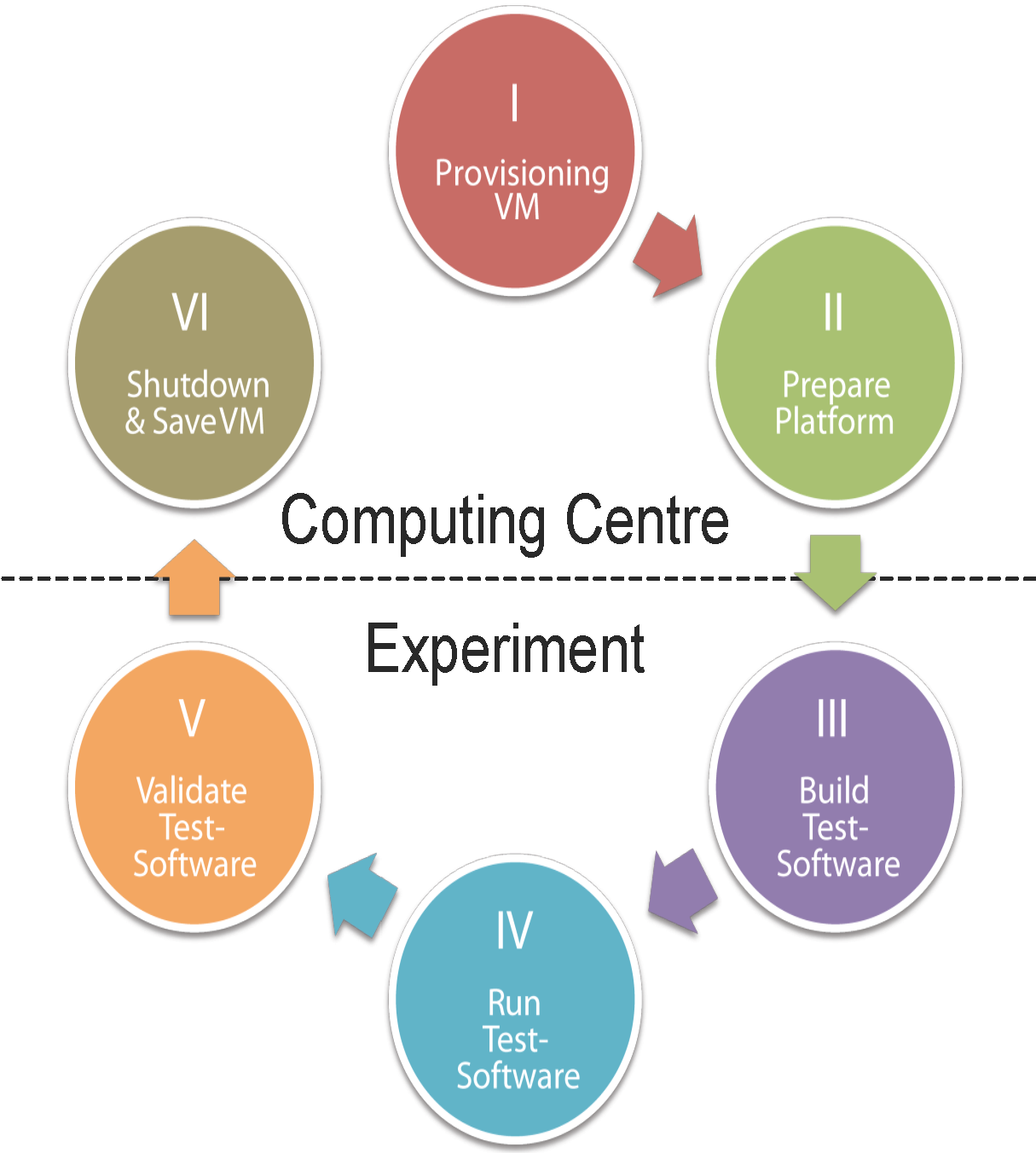}
  \vspace{-0.2cm}
  \caption{ \label{fig:validation} A sketch of the proposed software validation scheme under development at DESY.}
\end{wrapfigure}
In the case of HEP, data preservation means also preserving the
software and environment employed to analyse the data, equivalent to
level $3$ or $4$ of the preservation model.
Experience has shown that freezing the software in the latest state
would sustain analysis capability for only a limited amount of time.
Moreover, frozen simulation and reconstruction software would prevent
the implementation of improved reconstruction or calibration
algorithms or new MC simulations.
It is therefore beneficial to keep the experimental software running
and keeping it up--to--date with changes of the system architecture.
For this purpose it is advantageous to have a framework available to
automatically test and validate the software and data against changes
and upgrades to the environment, as well as changes to the
experimental software itself.
As such a framework would examine many facets common to several
current HEP experiments interested in a more complete data
preservation model, the development of a generic validation suite is
favourable.
A full version of such a suite is realised at DESY-IT, in co-operation
with the HERA experiments. The scheme, illustrated in
figure~\ref{fig:validation}, makes use of a virtual environment
capable of hosting an arbitrary number of different virtual machine
images and includes automated software build tools and data
validation.
Such a framework is by design expandable and able to host and validate
the requirements from multiple experiments.

\subsection{Global documentation initiatives}
\label{sec:inspire}
As well as the aforementioned individual documentation efforts, global
information infrastructures in HEP may be beneficial to the data
preservation project.
INSPIRE~\cite{inspirenet}, the successor to SPIRES, is an existing
third-party information system for HEP, and is thus ideally situated
to provide external management of experimental documentation.
The INSPIRE project is preparing for the ingestion of much more
high--level information in addition to the scientific papers
themselves.
These additions range from simple, documented information from the
experiments about a given analysis, through wikis and news--forums, to
even the data themselves, in a storage model where controlled access
is possible.
The additional information may, if desired, be only visible to
collaboration members, which allows internal information such as notes
to be stored, a project already in place between INSPIRE and the HERA
collaborations.
Another major advantage of such a scheme is that the responsibility of
hosting the information passes from a defunct experiment to an active
environment.

\subsection{HEP data for outreach, education and open access}
\label{sec:outreach}
The development of a HEP data format for outreach and education is an
attractive proposition.
For most collaborations such a project would run in parallel to
preserving the full re--analysis potential.
In recent years there is a notably increased global effort to improve
the overall level of education in particle physics and to provide
access to HEP to more people than ever before.
Tutorials using a simplified format of real HEP data with associated
pedagogical exercises might help further the public understanding of
science.
First projects have started within the BaBar~\cite{matt} and Belle~\cite{blab} 
collaborations.
The challenge of releasing such formats to the public domain is to 
keep the balance between a useful open access of HEP data on the one 
hand and control of the data, correctness and reputation of the experiment 
on the other hand.

\section{Resources, Governance and Structures}
\label{sec:governance}

The transition to a data preservation model should be planned in
advance of the anticipated end date of an experiment.
An early preparation is needed and sufficient resources should be
provided in order to maintain the capability to re--investigate older
data samples and collect all necessary documentation.
However, the additional resources are estimated to be rather small in
comparison to the person power allocated to the planning, construction
and running of an experiment.
Typically, a surge of $2$--$3$ FTEs for $2$--$3$ years, followed by
steady $0.5$--$1.0$ FTE per year per experiment is required for the
implementation of a level $3$ or $4$ preservation model, which should
be compared to $300$--$500$ FTEs per experiment for many years.
Therefore, the data preservation cost estimates represent typically
much less than $1$\% of the original investment, for a potential
$5$--$10$\% increase in physics output.
 

The transition to a new operational model of a collaboration should be
considered by HEP experiments before the end of their lifetime. If the
change to a long term analysis model is begun too late the
experimental collaboration risks being left in an undefined state.
In particular, the scientific supervision of the data preservation
process and decisions regarding authorship, access to data and
supervision of physics output after the collaboration's lifetime may
benefit from a restructuring of the collaboration towards the final
years.
The presence and influence of DPHEP may facilitate this transition by
providing common standards and global solutions within the HEP
community.

\section{Conclusions and Outlook}
\label{sec:outlook}

The collection of high energy physics data represents a significant
investment and physics cases can be made to demonstrate the potential
for scientific results beyond the lifetime of a collaboration.
However, until recently no coherent strategy existed regarding long
term access of HEP data and an international study group, DPHEP, was
formed to address this issue in a systematic way.
Given the current experimental situation, data preservation efforts in
HEP are timely, and large laboratories should define and install data
preservation projects in order to avoid catastrophic loss of data once
major collaborations come to an end.
The preservation of the full analysis capability of experiments,
including the reconstruction and simulation software, is recommended
in order to achieve a flexible and meaningful preservation model.
Such a project requires a strategy and well--identified resources, but
provides additional research at relatively low cost, enhancing the
return on the initial investment in the experimental facilities.


Data preservation efforts are best performed within the common
organisation at the international level DPHEP, through which there is
a unique opportunity to build a coherent structure for the future.
Common requirements on data preservation are now evolving via DPHEP
into inter--experimental R\&D projects, optimising the development
effort and potentially improving the degree of standardisation in HEP
computing in the longer term.
A second publication from the DPHEP group is expected shortly,
describing the current projects in more detail and providing
recommendations and guidelines for future HEP experiments.

\section*{References}
\bibliography{acat2011-kss}

\begin{figure}
\center
\vspace{-1cm}
\includegraphics[width=\textwidth]{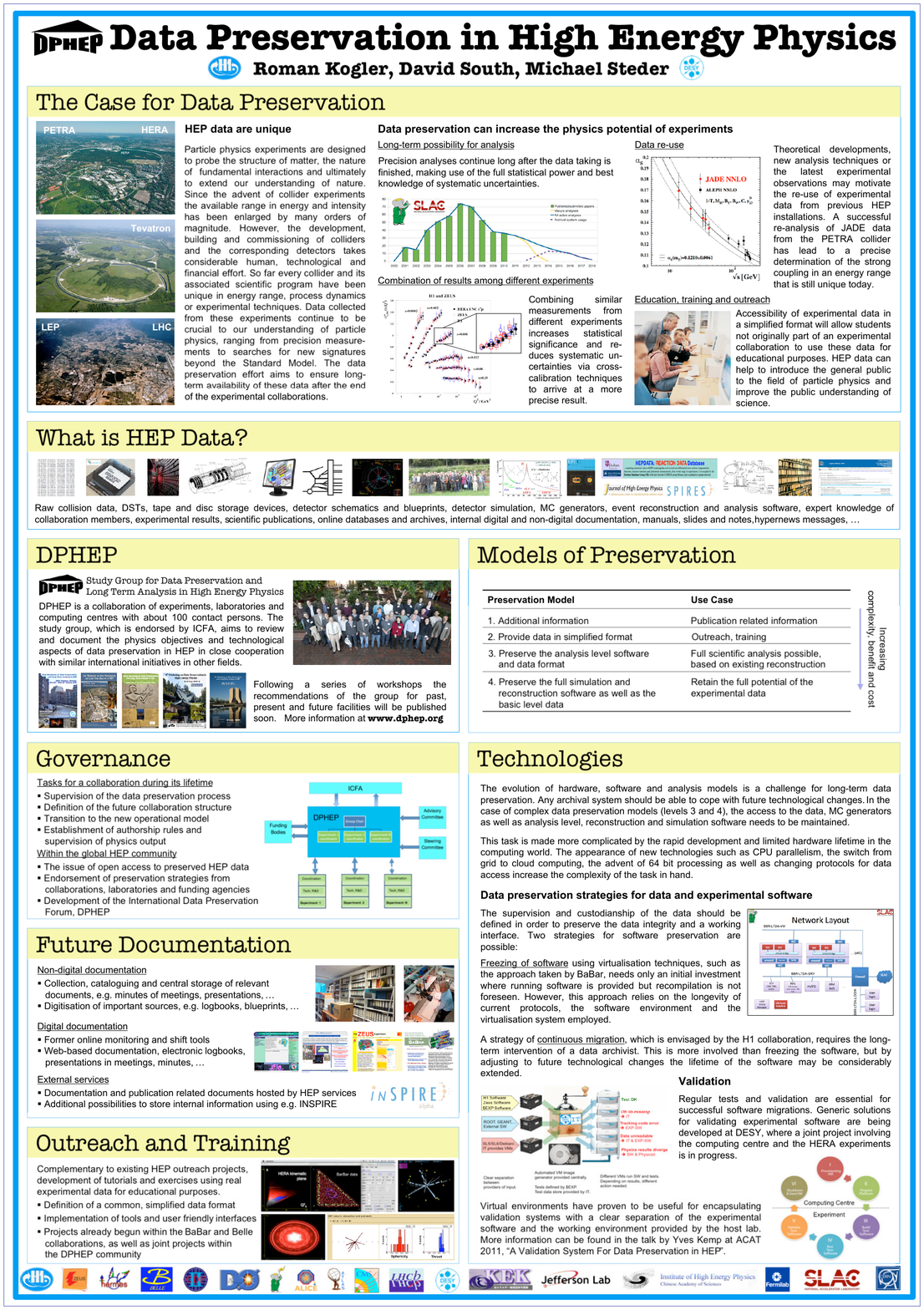}
 \caption{The poster presented at the ACAT2011 conference.}
\end{figure}

\end{document}